\documentstyle[epsf,twocolumn,floats,prb,aps]{revtex}
\begin{document}
\draft

\twocolumn[\hsize\textwidth\columnwidth\hsize\csname %
@twocolumnfalse\endcsname

\title
{Revisit Phase Separation of the Two-Dimensional $t-J$ model by the
Power-Lanczos Method}

\author{ C.T. Shih$^1$, Y.C. Chen$^2$, and T.K. Lee$^{3,4}$}
\address{
$^{1}$Division of Science Application, 
National Center for High-Performance Computing, Hsinchu, Taiwan\\
$^{2}$Dept. of Physics, Tunghai Univ., Taichung, Taiwan\\
$^{3}$
Inst. of Physics, Academia Sinica, Nankang, Taipei, Taiwan\\
$^{4}$National Center for Theoretical Sciences, Hsinchu, Taiwan
}

\date{\today}
\maketitle 
\begin{abstract} 
The power-Lanczos (PL) method is one kind of Green's function Monte-Carlo
simulation, which is improved by Lanczos iterations. The ground
state energies of strongly-correlated models can be evaluated
by this method quite accurately. In this report,
the boundary of phase separation (PS) of the two-dimensional $t-J$ model
is investigated by the power-Lanczos method and Maxwell construction.
The energies are compared with the results evaluated by other methods.
Our conclusion is that there is no phase separation for $J/t \le 0.4$.
\end{abstract}
                                                                               
\pacs{PACS numbers: 74.20.-z, 71.27.+a, 74.25.Dw}
]

\section{Introduction}
\label{s:intro}

It is believed
that several key features of the high-temperature superconductors (HTSC)
can be described by the two-dimensional (2D) $t-J$
model on square lattices. The Hamiltonian is:
\begin{equation}
H=-t\sum_{<i,j>\sigma} (\tilde{c}^+_{i\sigma}\tilde{c}_{j\sigma}+h.c.)
  + J\sum_{<i,j>}({\bf S}_i\cdot{\bf S}_j-{1\over4} n_in_j),
\label{e:tjm}
\end{equation}
where $<i,j>$ is the nearest-neighbor pairs and 
\begin{equation}
\tilde{c}_{i\sigma}=c_{i\sigma}(1-n_{i,-\sigma})
\label{e:constraint}
\end{equation}
One of the important questions on this model is whether the density
distribution of charge carriers in the CuO$_2$ planes is uniform
or not. There are several types of non-uniform phase such as
macroscopic phase separation which is separated into a hole-free
antiferromagnetic (AF) region and a hole-rich region\cite{emery90}, and
the stripe\cite{tranquada95,prelovsek93,white98,white98b}
phase which holes form domain walls separating AF regions.
The ground state is determined by the competition of
the kinetic ($t$) and exchange ($J$) terms. The $t$ term favors
the uniform phase to minimize the kinetic energy, while $J$ term
tend to attract the electrons together to lower the magnetic energy.

PS state and superconductivity are closely related. Some groups
argued that the driving mechanism of superconductivity is the
same as that of phase separation \cite{dagotto94} or
superconductivity comes from the frustrated phase separation
\cite{emery93}. On the other hand, if PS occurs in the physical
regime, the system will become insulating, and of course, no
superconductivity. Hence it is extremely important to determine the phase
separation boundary of the 2D $t-J$ model to resolve these issues.

Experimentally, phase separation of the superconducting
La$_2$CuO$_{4+\delta}$ compound are observed by neutron powder
diffraction \cite{jorgensen88}, $^{139}$La NMR \cite{hammel90,hammel91}
and magnetic susceptibility \cite{chou96}.
La$_2$CuO$_{4+\delta}$ phase separates for $0.01\leq\delta\leq0.06$
below T$_{ps}\approx300K$ into the nearly stoichiometric
antiferromagnetic La$_2$CuO$_{4+\delta_1}$ with $\delta_1$ less
than 0.02 and N\'eel temperature $T_N\approx250K$, and a metallic
superconducting oxygen-rich phase La$_2$CuO$_{4+\delta_2}$
with $\delta_2\approx0.06$ and superconducting transition
temperature $T_c\approx34K$.
In a recent transmission electron microscopy experiment,
an incommensurate modulation was observed directly in a
phase separated Cu-rich La$_2$CuO$_{4.003}$\cite{dong00}.
J. H. Cho {\it et al.} \cite{cho93}
measured the magnetic susceptibility of the Sr doped compound
La$_{2-x}$Sr$_x$CuO$_{4+\delta}$. For $x\leq0.03$, they also found
macroscopic phase separation into superconducting
La$_{2-x}$Sr$_x$CuO$_{4+\delta'}$ ($\delta'\approx0.08$) and
nonsuperconducting La$_{2-x}$Sr$_x$CuO$_{4+\delta''}$
($\delta''\approx 0.00$) phases. For $x\leq0.02$, the later phase is
antiferromagnetically ordered. The long range order disappears
and the phase becomes spin-glass-like for $0.02\leq x\leq 0.028$.
The doped holes in the AF (or spin-glass) phase
La$_{2-x}$Sr$_x$CuO$_{4+\delta''}$ ($0\leq x\leq 0.03$, $\delta''\approx 0.00$)
condense into walls separating microscopic undoped domains. The
phase becomes inhomogeneous electronically and magnetically.

On the other hand, the well known La$_{2-x}$Sr$_x$CuO$_4$
superconductor does not phase separate macroscopically because
the Sr$^{+2}$ ions are immobile and can not phase separate along
with the doped holes. In 1989, A. Weidinger {\it et al.}
\cite{weidinger89} observed coexistence of magnetic ordering
and superconductivity by muon spin rotation ($\mu^+$SR) experiments
performed on this material with $0\leq x\leq 0.15$.  Later D. R.
Harshman {\it et al.} \cite{harshman89} interpreted it as the
phase separation into an antiferromagnetic ordered phase and
a superconducting phase as in La$_2$CuO$_{4+\delta}$ discussed
above. J. H. Cho {\it et al.} \cite{cho92} measured the spin
dynamics of this material with $0.02\leq x\leq 0.08$ by the $^{139}$La
nuclear quadrupole resonance (NQR) spin-lattice relaxation rates
(NSLR) experiments in 1992. They concluded that the doped holes
were inhomogeneously distributed mesoscopically and segregated
into walls separating the hole-poor antiferromagnetic domains
with size $\approx a/\sqrt x\approx10-100\AA$. Later $\mu^+$SR
and NQR experiments also support this conclusion\cite{borsa94,borsa95}.
Recent NQR experiment shows that in single
crystal La$_2$CuO$_{4.02}$  there are three different regions with
different oxygen concentrations\cite{nikolaev00}.

The measurements of the superconducting diamagnetic moments
for underdoped and overdoped La$_{2-x}$Sr$_x$CuO$_4$ single
crystals show that the underdoped sample has only one
transition corresponding to $H_{c2}$, and the overdoped one
has two transitions with the higher one at $H_{c2}$.
The authors proposed that in the two-step transition comes
from the phase separation in the overdoped regime\cite{wen00}.

Several groups have studied the issue of PS
by analytical or numerical methods.
In the low electron density region, the theoretical results
studied by different methods are consistent.
Hellberg {\it et al.} determined very accurately
that the critical $J/t$ for phase separation at low
electron density limit is $J/t=3.4367$ \cite{hellberg95}.

But the theoretical 
results in the low-doping and small $J/t$ region are still conflicting. 
Emery {\it et al.}\cite{emery90} used the exact
diagonalization (ED) to study the $4\times4$ cluster. The 
Maxwell construction leads to the conclusion
that PS occurs for all values of $J/t$.
Hellberg and Manousakis (HM)\cite{hellberg97,hellberg00} investigated
this problem by the Green Function Monte Carlo (GFMC)
method and Maxwell construction
for larger clusters and reached the similar conclusion
that the $t-J$ model phase separates
for all values of $J/t$ in the low doping regime.
The $U(1)$ slave-boson functional integral results reach the
conclusion that for $0<J/t<0.2$ PS occurs for hole density
between $0$ and $0.1$\cite{tae00}.

However, the contradictory results are also claimed by several groups. 
Quantum Monte Carlo (QMC) \cite{moreo91,becca00} and
ED \cite{dagotto92} results on the Hubbard model, which should be consistent
with the $t-J$ model for small $J/t$, didn't give PS signal.
Putikka {\it et al.} studied this problem using the
high-temperature series expansion and found that at zero temperature,
PS does not occur for $J/t < 1.2$ at any electron density
\cite{putikka92,putikka00}. Prelov\v sek {\it et al.}\cite{prelovsek93} 
used ED to
study the two-point and four-point density correlations
on clusters of size 18 and 20 sites. They found that for $J/t>1.5$
the holes form domain walls along (1,0) or (0,1) direction,
and phase separate into a hole-rich and a
hole-free phase for even larger $J/t>2.5$. 
Poilblanc calculated the energy of 2 and 4
holes by ED on several clusters up to 26 sites.
The phase diagram includes a d-wave hole-pairing state
for $J/t \geq 0.2$, a liquid of four-hole droplets (quartets)
for larger $J/t \geq 0.5$, and at even larger $J/t$, an
instability towards PS \cite{poilblanc95}.
Yokoyama {\it et al.} investigated the phase diagram by
the variational  Monte Carlo (VMC) method \cite{yokoyama96}. The
critical $J/t$ for phase separation at the high density
limit they found is 1.5, which is consistent with Putikka
{\it et al.} 

The theoretical results of different groups discussed above
are consistent at the large $J/t$ and low electron density
region. But in the physical regime
of high T$_c$ superconductors,  $0.3<J/t<0.5$ and high electron
density $0.75<n_e<0.95$, they are conflicting with each other.
We have used the power-Lanczos 
(PL) method \cite{yctk95,heeb93}
to obtain the best estimate of the ground state
energy in this physical regime 
for the largest cluster (122 sites) that have been studied 
so far. The same lattices and boundary conditions as those
used by HM \cite{hellberg00} are studied
also in this report. It can be shown that our data {\it
calculated} by the PL method, which is a rigorous upper
bound of the ground state energies, 
are still lower than {\it extrapolated} GFMC data reported
by HM in the physically interested
regime. Based on the variational argument we show
that there is no phase separation in this physical regime\cite{khono97}.
In this report, we will focus on the competition of the uniform state
and PS state. The stripe phase is not taken into account.

\section{Numerical Method}
\label{s:method}

The numerical method used here is the ``power-Lanczos method'',
which is a revised version of Green Function Monte Carlo (GFMC)
method\cite{linden92}, and improved by the Lanczos iterations such that
the ground state properties can be calculated more exactly.

The trial wave functions we used in this report is the
d-wave RVB (resonating valence bond) with AF
correlation \cite{tk88},
\begin{equation}
\mid \Psi^T\rangle = exp(hS_M)P_d\prod_k
   (u_k+v_kc^\dagger_{k\uparrow}c^\dagger_{-k\downarrow}|0\rangle
\label{e:twf}
\end{equation}
with
\begin{eqnarray}
u_k^2&=&\frac{1}{2}(1-\frac{\epsilon_k}{E_k}), \;
v_k^2=\frac{1}{2}(1+\frac{\epsilon_k}{E_k}) \nonumber\\
\epsilon_k&=&-2(cosk_x+cosk_y)-\mu, \;
E_k=\sqrt{(\epsilon_k^2+\Delta_k^2)}
\label{e:rvb}
\end{eqnarray}
where $\Delta_k$ and $h$ are variational parameters controlling the
magnitude of superconductivity and antiferromagnetism. 
$S_M=\sum_i (-1)^{x_i+y_i} S_z^i$, which is the staggered
magnetization.
$P_d$ operator excludes the states with doubly occupied sites
from the wave function to satisfy the constraint of Eq.(\ref{e:constraint}).
By tuning
the two parameters, the variational energy can be minimized
and the trial wave function corresponding to the best
set of parameters will be the starting point of the power-Lanczos
projection.

The ``power method'' is the simplified Green
function Monte Carlo (GFMC) method, which projects
the trial wave function determined variationally toward the true
ground state wave function of the Hamiltonian systematically.

Suppose the optimal trial wave function is $\mid \Psi^T \rangle$.
The eigenstates of Hamiltonian are $\mid \Psi^i \rangle$ and
$E_i$ are the corresponding eigenenergies.
$\mid \Psi^0 \rangle$ denotes the ground state wave function.
$\{\mid \Psi^i \rangle\}$ form a complete set of wave functions
and $\mid \Psi^T \rangle$ can be expanded as
\begin{equation}
\mid \Psi^T \rangle = \sum_i a_i \mid \Psi^i \rangle
\label{e:pwm1}
\end{equation}
Applying the operator $(W-H)^p$ on $\mid \Psi^T \rangle$, we get
\begin{eqnarray}
\mid \Psi^{(p)} \rangle &=& (W-H)^p \mid \Psi^T \rangle
   = \sum_i (W-E_i)^p a_i \mid \Psi^i \rangle\nonumber\\
   &=& (W-E_0)^p \sum_i (\frac{W-E_i}{W-E_0})^p a_i
\mid \Psi^i \rangle
\label{e:pwm2}
\end{eqnarray}
where $W$ is some constant. If $W$ is properly chosen such that
\begin{equation}
\mid W-E_0 \mid > \mid W-E_i \mid
\label{e:pwm3}
\end{equation}
for all $i \ne 0$, the weight of excited states will become
smaller and smaller when the power $p$ is increasing. In the
$p \to \infty$ limit, all $(\frac{W-E_{i\ne 0}}{W-E_0})^p$
become zero and $\mid \Psi^{(p)} \rangle \to \mid \Psi^0
\rangle$. The trial wave function is projected to the ground
state by applying infinite powers of $(W-H)$. For the $t-J$
model, Eq.(\ref{e:pwm3}) can be satisfied by setting $W=0$. 
If $\mid \Psi^T \rangle$ has the same quantum numbers (${\bf S}$,
$S^z$, ${\bf k}$, etc.) as the ground state $\mid \Psi^0 \rangle$,
and $\mid \Psi^T \rangle$ is not orthogonal to $\mid \Psi^0 \rangle$,
i.e., $a_0\ne 0$,
$\mid \Psi^{(p)}\rangle=H^p\mid \Psi^T
\rangle$ will approach to the ground state as $p \to \infty$.
For a physical quantity $O$ the expectation value of power $p$
is 
\begin{eqnarray}
\langle \Psi^{(p)}\mid O \mid \Psi^{(p)}\rangle/
\langle \Psi^{(p)}\mid\Psi^{(p)}\rangle = \nonumber\\
\langle
\Psi^T\mid H^p O H^p\mid \Psi^T\rangle / \langle \Psi^T\mid
H^{2p} \mid \Psi^T \rangle
\end{eqnarray}

Theoretically we can always get the ground state by the power
method if we have the trial wave function with correct quantum
numbers. However, there are two difficulties with this method.
First, the number of intermediate states grow exponentially
with powers.
Second, the statistical
error bar grows exponentially with the applied power due to
the famous `sign' problem of the Fermionic system. We will
discuss these two problems below.

A configuration $\mid a \rangle$ in real space is defined
as
\begin{equation}
\mid a \rangle = \prod^{N_e/2}_{i=1} c^\dagger_{r_i\uparrow}
\prod^{N_e/2}_{j=1} c^\dagger_{r_j\downarrow}\mid 0\rangle
\end{equation}
where $r_i$ and $r_j$ represent the positions
of the i$th$ up and j$th$ down spins, respectively.
$|a\rangle$ satisfies
\begin{equation}
H\mid a \rangle = \sum_{a'} \mid a'\rangle \langle a' \mid
H\mid a \rangle
\end{equation}
From Eq.(\ref{e:tjm}), we see that there are three kinds of
$\mid a'\rangle$ such that the coefficient $\langle a'\mid H
\mid a\rangle \ne 0$: flipping a nearest-neighbor pair of spin
up and down, moving an electron from an occupied site to an
empty site, and $|a\rangle$ itself. The number of such
$|a'\rangle$ is of order $N_e$, the number of electrons. 
When we apply the next $H$
on it, $H|a'\rangle$ will become a summation of about
order $N_e$ terms again. To measure the expectation value
of some physical quantity $O$ to the $p$th power,
there are $\sim N_e^{2p}$ terms
to be calculated. So it is impossible to calculate physical
quantities for large power $p$.
To avoid this difficulty, we will not sum all the $O(N_e^{2p})$
intermediate states for the chosen configurations. Only the
important part of
them are taken into account by using Monte Carlo method.

The first step is to generate the sequence of $N_{total}$
configurations $\mid a_i \rangle$, $1 \le i \le N_{total}$, by
Metropolis algorithm which is distributed as the weight
$P(i)$ corresponding to the trial wave function
optimized by VMC method\cite{ceperley77}.
Our task now is to calculate the
matrix elements
\begin{eqnarray}
\langle O^{(p)} \rangle_0 &=& \frac{\langle \Psi^T \mid H^pOH^p
 \mid \Psi^T \rangle}
{\langle \Psi^T \mid \Psi^T \rangle} = \sum_i P(i)O^{(p)}(i)\nonumber\\
P(i)&=&\frac{\mid C_i \mid^2}{\sum_{i'} \mid C_{i'} \mid^2}
\nonumber\\
O^{(p)}(i)&=&\frac{\sum_{i'} \langle a_i \mid H^pOH^p \mid a_{i'}
\rangle C_{i'}}{C_i}
\label{e:pwm4}
\end{eqnarray}
where $C_i$ is the amplitude of the configuration $\mid a_i\rangle$
of the trial wave function $\Psi^T\rangle$.
The operator $O$ equals to the Hamiltonian operator $H$ when
we measure the energy.
The first task is to calculate $\langle H^n \rangle$ which
will be useful for calculating energy ($n=2p+1$), and normalization
factor ($n=2p$). Expanding the matrix element in the following form
\begin{eqnarray}
\langle a_i \mid H^n \mid a_f \rangle &=&
\sum_{m_1,m_2,\ldots,m_{n-1}} \langle a_i \mid H \mid m_1 \rangle
\langle m_1 H \mid m_2\rangle\ldots\nonumber\\
&& \langle m_{n-2}\mid H \mid m_{n-1}
\rangle \langle m_{n-1} \mid H \mid a_f\rangle \label{e:all_inter}
\end{eqnarray}
Define
\begin{equation}
P_{\alpha\beta}=P_{|\alpha\rangle \to |\beta\rangle} =
\frac{1}{Z_\alpha} H_{\alpha\beta}
\label{e:rmwk_p}
\end{equation}
with $Z_\alpha = \sum{}_\beta H_{\alpha\beta}$ and $H_{\alpha
\beta} = \langle \alpha \mid H \mid \beta \rangle$.
Eq.(\ref{e:all_inter}) can be rewritten as
\begin{eqnarray}
\langle a_i \mid H^n \mid a_f \rangle &=&
\sum_{m_1,m_2,\ldots,m_{n-1}} (P_{a_i m_1} Z_{a_i})
(P_{m_1 m_2} Z_{m_1})\ldots\nonumber\\
&&(P_{m_{n-1} a_f} Z_{m_{n-1}})
\label{e:rmwk}
\end{eqnarray}
To calculate this exactly all the possible paths $\{m_1, m_2,
\ldots, m_{n-1}\}$ connecting the initial and final configuration
$\mid a_i \rangle$ and $\mid a_f \rangle$ have to be taken into
account. Since there are about $N_e^n$ terms, it
is impossible to sum all the terms. Instead we make use of the
random walk procedure.

Starting from the initial configuration $\mid a_i \rangle$,
use $P_{a_i m'_1}$ defined in Eq.(\ref{e:rmwk_p}) to be the
probability of transition from $\mid a_i \rangle$ to $\mid
m'_1\rangle$ state. There are many possible $\mid m'_i \rangle$
of three types
corresponding to nearest neighbor spin-flipping, hole-hopping,
and diagonal terms as mentioned earlier in this section.
For these types the corresponding matrix elements $H_{a_im'_1}$
are $-t$, $-J/2$, and $-J/2$, respectively. Choose one of these
$\mid m'_1\rangle$ to be the $\mid m_1 \rangle$ randomly with the
transition probability $P_{a_i m'_1}$. Then use the same procedure
to generate the next intermediate state $\mid m_2 \rangle$. Repeat
this $n$ times, we can arrive the final state $\mid a_f\rangle$ of
this path. The measurement of $\langle H^n \rangle$ for this
path is
\begin{equation}
M_i=(Z_{a_i}\cdot Z_{m_1}\cdot Z_{m_2}\cdots Z_{m_{n-1}})
C^T_{a_f}/C^T_{a_i}
\label{e:1walker}
\end{equation}
This is only one term in the summation of Eq.(\ref{e:rmwk}).
Of course the error will be large if we use this value to estimate
the summation in Eq.(\ref{e:rmwk}). To reduce the error, we
generate many paths by the procedure above and average the
measurement
\begin{equation}
\bar{M_i}=\frac{1}{N_{path}}\sum_{k=1}^{N_{path}}
  Z_{a_i}Z_{m^k_1}Z_{m^k_2}Z_{m^k_3}\cdots Z_{m^k_{n-1}}
  \cdot\frac{C^T_{a^k_f}}{C^T_{a_i}}
\label{e:pwm5}
\end{equation}
$N_{path}$ is total number of paths of random walk
with the same initial configuration $|a_i\rangle$.
$\{|m^k_1\rangle,
|m^k_2,\ldots,|a^k_f\rangle\}$ is the intermediate and final states
of the $k$th path. 
$N_{path}$ should be large enough to get good
statistics. Typical $N_{path}$
is of order $10^2 \sim 10^3$. Together with Eq.(\ref{e:pwm4}),
and Eq.(\ref{e:pwm5}) we get
\begin{equation}
\langle H^{n} \rangle = \langle M\rangle = \frac{1}{N_{total}}
\sum_{i=1}^{N_{total}} \bar{M_i} \pm \delta
\label{e:pwm_eng}
\end{equation}
with
\begin{eqnarray}
\delta &=& \sqrt{\frac{1}{N_{total}}
(\langle M^2\rangle -\langle M\rangle^2)}\nonumber\\
\langle M^2\rangle &=& \frac{1}{N_{total}}
\sum_{i=1}^{N_{total}} \bar{M_i}^2
\end{eqnarray}
Now we can calculate $\langle H^n \rangle$. With this it is
straightforward to calculate the energy of $p$th power $E^{(p)}
=\langle H^{2p+1} \rangle / \langle H^{2p} \rangle$.

Another difficulty is the `fermionic sign problem'. The matrix
elements of Hamiltionian $\langle \alpha\mid H\mid\beta\rangle$
are not positive or negative definite. That is, some of
the weight Eq.(\ref{e:1walker}) of the $N_{path}$ paths are
positive and others are negative. The summation in Eq.(\ref{e:pwm5}),
or Eq.(\ref{e:pwm_eng}) will become a
summation of a positive part and a negative part
\begin{equation}
\langle H^p O H^p \rangle = (O^{(p)}_+ + O^{(p)}_-) \pm
(\sigma^{(p)}_+ + \sigma^{(p)}_-)
\label{e:sign}
\end{equation}
where $O^{(p)}_+$ and $O^{(p)}_-$ are the positive and negative
parts of the summation in Eq.(\ref{e:pwm_eng}).

\begin{figure}[htb]
\epsfysize=9cm
\epsfbox{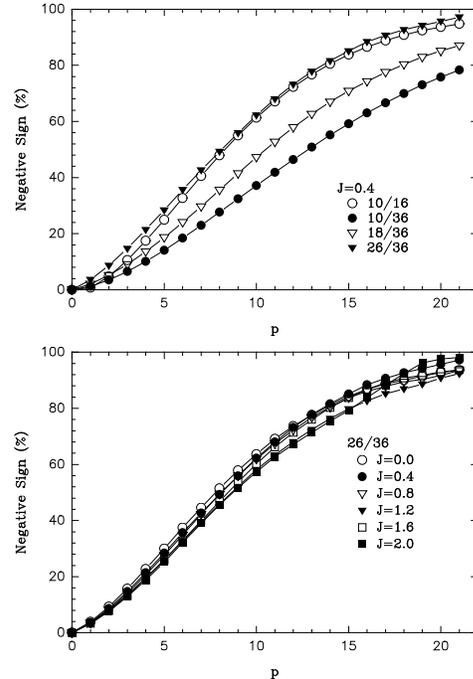}
\caption{
Negative sign ratio which is defined as the ratio of the absolute value
of negative and positive parts of $\langle H^n\rangle$ for (a) different
electron densities: 10/16 (open circle), 10/36 (full circle), 18/36
(open triangle) and 26/36 (full triangle) and (b) different $J/t$
values: $J/t=0$ (open circle), 0.4 (full circle), 0.8 (open triangle),
1.2 (full triangle), 1.6 (open square) and 2.0 (full square).}
\label{f:l0sign}
\end{figure}

For smaller powers, the portion of
positive part is much larger than the negative part. It is
because that a step of random walk only flips a pair of nearest
neighbor spins or moves an electron to the nearest neighbor
empty site, and the probability of crossing the nodes of the trial
wave function and resulting a negative matrix element is small.
For large powers, the probability of crossing the nodes of the trial
wave function will be accumulated during the long path of random
walk and the negative part in the summation Eq.(\ref{e:pwm5}) and
Eq.(\ref{e:pwm_eng}) will grow larger. The larger the negative
portion (absolute value closer to the positive portion) the larger
the error bars of the physical measurements will be. In
Fig.\ref{f:l0sign}(a) we show the ratio of the negative and
positive portions varies with power for $J/t=0.4$ and electron
density $n_e=10/16$, $10/36$, $18/36$, and $26/36$. It is clear
that the negative portion is larger for higher electron density.
For the high density cases $10/16$ and $26/36$, the absolute value
of negative part is almost the same as the positive part for
$\langle H^{20} \rangle$. Thus $\mid \langle H^{20}\rangle\mid
\ll \mid\langle H^{20}\rangle_\pm\mid$ and the statistical error
of $\langle H^{20}\rangle$ will be largely increased. For example,
the negative ratio is about 79\% for 10/36, 87\% for
18/36, and 97\% for 26/36.

\begin{figure}[htb]
\epsfysize=12cm\epsfbox{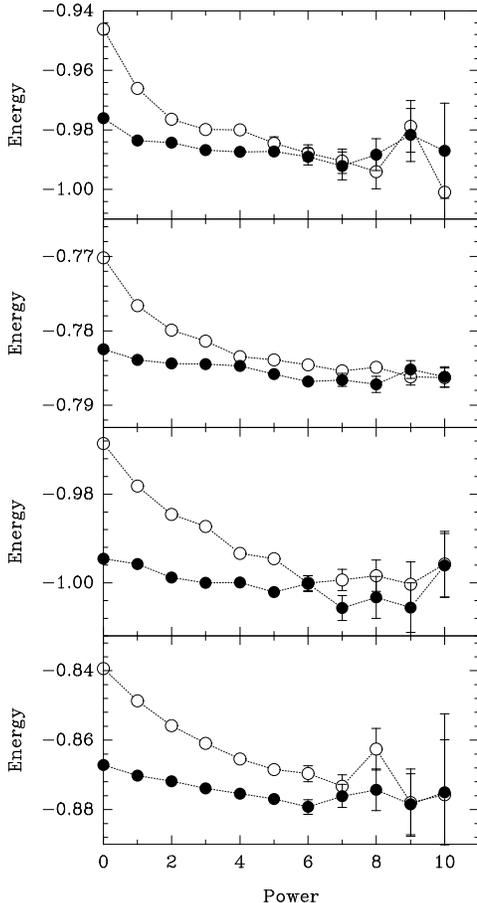}
\caption{
Energy vs power for $J/t=0.4$ for 10/16, 10/36, 18/36,
and 26/36 (from the top to the bottom, open circles). The full circles
are the PL1 data.}
\label{f:l1_eng}
\end{figure}

If we choose $N_{total}$ and $N_{path}$ large enough to reduce
the error bars of $\langle H^{20}\rangle_\pm$ to be as small as
0.1\%, the corresponding error bars of $\langle H^{20}\rangle$
of these three electron densities will be 1\%, 1.5\%, and 6.7\%.
So no matter how large $N_{total}$ and $N_{path}$ are, and how
accurate $O^{(p)}_\pm$ are, the sign problem always prevents us
to do calculations for large powers, especially in the high
electron density regime. The largest power we use is usually
6 or 8 for high electron densities. The negative ratio is about
80\%$\sim$90\%. Beyond this the statistical errors are
too large to be reliable. The energies varies with power
for different densities are plotted in the
Fig.\ref{f:l1_eng} (open circles) for $J/t=0.4$. 
In Fig.\ref{f:l0sign}(b) the negative ratio
for different $J/t$ value of 26/36 system is shown. By comparing
Fig.\ref{f:l0sign}(a) and (b) we see the negative ratio depends
on the electron density much more sensitive than $J/t$.

The sign problem is the intrinsic limitation of the power method.
It is almost impossible to project the trial wave function to be very
close to the ground state by applying large powers on it, especially
in the physical interested region with high electron density.
Other methods like QMC or GFMC also encounter a similar problem.
To overcome this limitation, we have modified the power method.
Here we will introduce the more powerful power-Lanczos
method\cite{yctk95}.

An inappropriately chosen trial wave function would require a lot
of computer time to converge to the ground state because it is
necessary to apply many powers on the trial wave function.
Hence large $N_{total}$ and $N_{path}$ will be necessary
to reduce the statistical error due to the random walk
approximation and the fermionic negative sign problem.
Since the negative sign problem prevents us to do large powers to
get the results close to the ground state in the physically
interested high density region, another solution is to make our
starting point, the trial wave function closer to
the ground state. If our trial wave function
is closer to the ground state, it will not be necessary to apply
many powers. And the negative impact
of sign problem can be reduced. Here we use the Lanczos algorithm
to improve our trial wave function, and then apply powers of
Hamiltonian on the new trial wave function. The method is a
combination of power method introduced in the previous section
and the Lanczos iteration method usually used in the exact
diagonalization (ED) for the small clusters. Hence it was named as
`power-Lanczos method'. 
Lanczos iteration method was first proposed by Heeb and Rice
to improve the trial wave function
systematically \cite{heeb93}. The effectiveness of the method
is demonstrated on the two-dimensional antiferromagnetic
Heisenberg model and later on the two-dimensional $t-J$ model
\cite{heeb94}. We will describe their method below.

Starting from a trial wave function $\mid \phi_0\rangle$
(which we called it $\mid\Psi^T\rangle$ above)
optimized by the variational method.
Apply the Hamiltonian $H$ on $\mid \phi_0\rangle$. Since
$\mid \phi_0\rangle$ is not an eigenstate of $H$, it becomes
\begin{equation}
H\mid \phi_0\rangle = a_0\mid \phi_0\rangle + b_1\mid \phi_1\rangle
\label{e:Htwf}
\end{equation}
with
\begin{equation}
\langle \phi_0\mid\phi_1\rangle = 0
\end{equation}
which means $\mid \phi_0\rangle$ and $\mid \phi_1\rangle$ are
orthogonal. $a_0=\langle\phi_0\mid H\mid\phi_0\rangle=E_0$, which
is the expectation value of energy of the trial wave function. And
\begin{eqnarray}
b_1 &=& \langle\phi_1\mid H\mid\phi_0\rangle\nonumber\\
  &=& \frac{1}{b_1}\langle\phi_0\mid (H-E_0)H\mid\phi_0\rangle
\label{e:pl_a}
\end{eqnarray}
This formalism is similar to the Lanczos iteration
for exact diagonalization, which the recurrence relation is
\begin{eqnarray}
H\mid\phi_0\rangle &=& a_0\mid\phi_0\rangle + b_1\mid\phi_1\rangle
\nonumber\\
H\mid\phi_n\rangle &=& a_n\mid\phi_n\rangle + b_n\mid\phi_{n-1}
\rangle + b_{n+1}\mid\phi_{n+1}\rangle
\label{e:lanczos}
\end{eqnarray}
For small systems, $n$ can be as large as the number of all the
possible configurations in real space and $H$ can be diagonalized exactly
in the basis $\{\mid\phi_0\rangle,\mid\phi_1\rangle,\ldots,
\mid\phi_n\rangle\}$ and the eigenstate corresponding the lowest
eigenenergy is just the ground state we need.
The largest cluster solved by this method
exactly is 32 sites\cite{leung95}. Although for larger clusters it is
impossible to use this method to solve the exact ground state,
we can truncate the recurrence relation Eq.(\ref{e:lanczos}) at n
which is small enough such that we can calculate at large systems,
and then diagonalize the Hamiltonian by the incomplete basis.
The lowest-energy eigenstate will not be the ground state because
the basis is incomplete but it will be much closer to the ground state
than the original trial wave function even $n$ is small.

Now consider the first order iteration ($n=1$) which we call it
`PL1' in Eq.(\ref{e:lanczos})
\begin{eqnarray}
H\mid\phi_0\rangle &=& a_0\mid\phi_0\rangle + b_1\mid\phi_1\rangle
\nonumber\\
H\mid\phi_1\rangle &=& a_1\mid\phi_1\rangle + b_1\mid\phi_0
\rangle + b_2\mid\phi_2\rangle
\label{e:pl_1}
\end{eqnarray}
where $\langle\phi_2\mid\phi_0\rangle=0$ and $\langle\phi_2\mid
\phi_1\rangle=0$. Since we truncate the recurrence relation
Eq.(\ref{e:lanczos}) at $n=1$ here, the last term in Eq.(\ref{e:pl_1})
is dropped and the Hamiltonian can be written in the space spanned
by the basis $\{\mid\phi_0\rangle,\mid\phi_1\rangle\}$ as
\begin{equation}
H = \left(
\begin{array}{cc}
H_{00} & H_{01}\\
H_{10} & H_{11}
\end{array}
\right) = \left(\begin{array}{cc}
E_0 & b_1\\
b_1 & a_1\end{array}\right)
\label{e:Hpl1}
\end{equation}
where $E_0=a_0$.
Remember that this is the first order approximation because we
drop the second order term in Eq.(\ref{e:pl_1}). Under this
approximation, $a_1$ in Eq.(\ref{e:pl_1}) is
\begin{eqnarray}
a_1 &=& \langle \phi_1\mid H\mid\phi_1\rangle\nonumber\\
    &=& \frac{1}{b_1^2}\langle \phi_0\mid (H-E_0)H(H-E_0)\mid
    \phi_0\rangle \nonumber\\
  &=& \frac{\langle\phi_0\mid (H-E_0)H(H-E_0)\mid\phi_0\rangle}
     {\langle\phi_0\mid (H-E_0)H\mid\phi_0\rangle}
\label{e:pl_b}
\end{eqnarray}
Diagonalize the
Hamiltonian in Eq.(\ref{e:Hpl1}), the smaller eigenvalue and the
corresponding eigenstate is
\begin{eqnarray}
E_1 &=& \frac12[E_0+a_1-\sqrt{(E_0-a_1)^2 + b_1^2}]\nonumber\\
\mid PL1\rangle &=& \mid\phi_0\rangle + \frac{b_1}{E_1-a_1}
\mid\phi_1\rangle
\label{e:pl1twf}
\end{eqnarray}
We have to know the weight of each configuration in real space
of the new trial wave function $\mid PL1\rangle$. If
\begin{equation}
\mid PL0\rangle = \mid\phi_0\rangle =\sum_i a^{(0)}_i\mid i\rangle
\nonumber
\end{equation}
and
\begin{equation}
\mid PL1\rangle =\sum_i a^{(1)}_i\mid i\rangle
\end{equation}
where $a_i^{(0)}$ is the amplitude of $\mid i\rangle$. Then
\begin{eqnarray}
a^{(1)}_i &=& \langle i\mid PL1\rangle\nonumber\\
  &=& \langle i\mid(\mid PL0\rangle
+ \frac{b_1}{E_1-a_1}\mid\phi_1\rangle)\nonumber\\
  &=& a^{0}_i +\frac{b_1}{E_1-a_1}
\langle i\mid\phi_1\rangle
\nonumber\end{eqnarray}
From Eq.(\ref{e:Htwf}) we know $b_1\mid\phi_1\rangle=(H-E_0)\mid
PL0\rangle$. Thus
\begin{eqnarray}
a^{(1)}_i &=& a^{(0)}_i + \frac{1}{E_1-a_1}\langle i\mid (H-E_0)
\mid PL0\rangle\nonumber\\
  &=& a^{(0)}_i(1-\frac{E_0}{E_1-a_1})+\frac{1}{E_1-a_1}
      \sum_j \langle i\mid H\mid j\rangle a^{(0)}_j\nonumber
\end{eqnarray}
Thus the ratio of the weight of $\mid PL1\rangle$ and
$\mid PL0\rangle$ corresponding the same configuration
$\mid i\rangle$ is
\begin{equation}
\frac{a^{(1)}_i}{a^{(0)}_i} = 1-\frac{E_0}{E_1-a_1} +
    \frac{1}{E_1-a_1}\sum_j\langle i\mid H\mid j\rangle
    \frac{a^{(0)}_j}{a^{(0)}_i}
\label{e:pl_2}
\end{equation}
Heeb and Rice \cite{heeb93} proposed to calculate $a_1$ and $b_1$
by using Monte Carlo technique. With $a_1$ and $b_1$, $E_1$ and
$\mid PL1\rangle$ can be calculated from Eq.(\ref{e:pl1twf}).
However, in this method $a_1$ and $b_1$ have to be calculated
very accurately. A small error will produce large uncertainty
in $E_1$ and $\mid PL1\rangle$. The statistical error is inevitable since
they are calculated by Monte Carlo method. Here we choose an
alternative. Since $\mid PL1\rangle$ in Eq.(\ref{e:pl1twf})
has not been properly normalized. All the $a^{(1)}_i$ can be
divided by a constant $(1-\frac{E_0}{E_1-a_1})$ and
Eq.(\ref{e:pl_2}) becomes
\begin{equation}
\frac{a^{(1)}_i}{a^{(0)}_i} = 1 + C_1
  \sum_j \langle i\mid H\mid j\rangle\frac{a^{(0)}_j}{a^{(0)}_i}
\label{e:pl1wgt}
\end{equation}
with
\begin{equation}
C_1 = \frac{1}{E_1-a_1-E_0}
\end{equation}

Note that $C_1$ is a constant once the trial wave function
$\mid PL0\rangle = \mid\phi_0\rangle$ is chosen. Instead of
calculating $C_1$ by using $a_1$ and $b_1$ obtained by
Monte Carlo method, we treat $C_1$ as a variational
parameter to optimize the energy. The optimized energy is 
just $E_1$. For each configuration the PL0
weight of the trial wave function has to be calculated first
and the PL1 weight can be evaluated by using Eq.(\ref{e:pl1wgt}).
This procedure largely increases the computer time (approximately
of order $N_e$). To save computer time Ceperley's algorithm for ratio of
weight\cite{ceperley77} is used for calculating
$a^{(0)}_j/a^{(0)}_i$ where $\mid i\rangle$ and $\mid j\rangle$
are connected by $H$. It is important to
check the ratio of determinant to prevent the divergence of
the ratio.
$a^{(0)}_j$ and $a^{(0)}_i$ should be calculated
directly if their ratio obtained by Ceperley's method is too
large or too small, for example, larger than $10^8$ or less than
$10^{-8}$. Otherwise the results evaluated by using Ceperley's
method might be in err.

\begin{figure}[htb]
\epsfysize=7cm\epsfbox{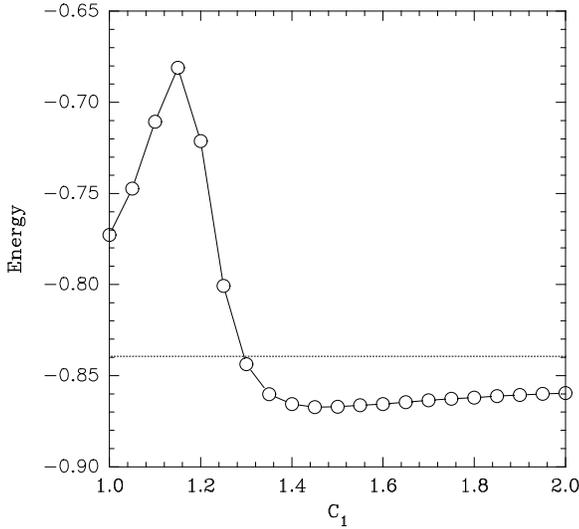}
\caption{PL1$_{power=0}$ energy dependence of $C_1$ for 26/36, $J/t=0.4$.
the horizontal dotted line represents the energy of PL0$_{power=0}$.}
\label{f:pl1c1}
\end{figure}

To estimate the possible range of $C_1=1/(E_1-a_1-E_0)$, from
Eq.(\ref{e:pl_b}) it is easy to show
\begin{equation}
a_1 = \frac{\langle\phi_0\mid (H-E_0)^3\mid\phi_0\rangle}
         {\langle\phi_0\mid (H-E_0)^2\mid\phi_0\rangle}+E_0
\nonumber
\end{equation}
If our trial wave function is not too bad, roughly we can estimate
that $E_1 \approx E_0$ and $a_1 \approx E_0$. So the range we tuning
$C_1$ is around $-1/E_0$. Typical behavior of $E_1$ vs $C_1$ is shown
in Fig.\ref{f:pl1c1}. 

From our experience of PL1 calculation, the PL1$_{power=0}$ energy $E_1$
is approximately equal to the energy of PL0$_{power=4}$ or
PL0$_{power=5}$. Our new trial wave function $\mid PL1\rangle$ is
much better than the original trial wave function $\mid PL0\rangle$.
The next step is to improve the wave function further. There are
two ways to achieve this purpose: (1) using power method introduced
in the previous section, applying powers on $\mid PL1\rangle$ and
projecting it to the ground state systematically; (2) using the
next order of Lanczos iteration, i.e., the `PL2'.

\begin{figure}[htb]
\epsfysize=7cm
\epsfbox{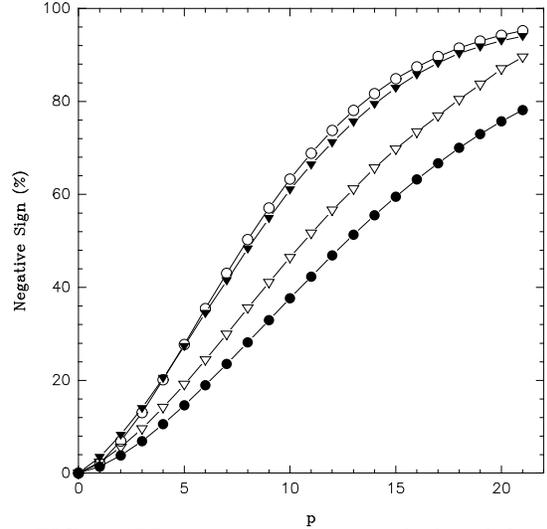}
\caption{
PL1 negative sign ratio which is defined as the ratio of the absolute value
of negative and positive parts of $\langle H^n\rangle$ for different
electron densities: 10/16 (open circle), 10/36 (full circle), 18/36
(open triangle) and 26/36 (full triangle).}
\label{f:l1sign}
\end{figure}

First let's see the first method: `PL1-power' calculation.
We can use the same procedure introduced previously
to apply $H^p$ on the new wave function $\mid PL1\rangle$.
The only difference is that the weight of the configurations
has to be changed from PL0 to PL1 (Eq.(\ref{e:pl1wgt})). As in
the PL0-power calculation, the PL1-power calculation is suffered
from the fermionic sign problem. In Fig.\ref{f:l1sign} the
negative sign ratio vs power for $\mid PL1\rangle$
is shown for $J/t=0.4$ in 10/16, 10/36, 18/36, and 26/36 lattices.
It is clear that the behavior of negative sign ratio vs power
for $\mid PL0\rangle$ and $\mid PL1\rangle$ are almost the same,
which means the maximum power under the limitation of negative
sign problem is the same for both $\mid PL0\rangle$ and
$\mid PL1\rangle$. Since $\mid PL1\rangle$ is closer to the
ground state than $\mid PL0\rangle$, the best PL1-power energy
will be lower than that of PL0. Fig.\ref{f:l1_eng} shows the
PL0 and PL1-power energy for $J/t=0.4$ in 10/16, 10/36, 18/36, and
26/36 sites (full circles). 
It is clear that PL1 always yields lower energies.

Another way to improve the wave function further is to do PL2.
Since we have the equations for $\mid PL1\rangle$, we can follow
the same procedure which evaluates $\mid PL1\rangle$ from
$\mid PL0\rangle$ to generate $\mid PL2\rangle$ from
$\mid PL1\rangle$. For a configuration $\mid i\rangle$ the
weight corresponding $\mid PL2\rangle$ is
\begin{equation}
a^{(2)}_i = a^{(1)}_i + C_2\sum_j\langle i\mid H\mid j\rangle
  a^{(1)}_j\nonumber
\end{equation}
which is completely in the same form as Eq.(\ref{e:pl1wgt}).

Using Eq.(\ref{e:pl1wgt}) to replace $a^{(1)}_i$ by the expression
of $a^{(0)}_i$ we get
\begin{eqnarray}
a^{(2)}_i &=& a^{(0)}_i + C_1\sum_k\langle i\mid H\mid
  k\rangle a^{(0)}_k \nonumber\\
  &+& C_2\sum_j\langle i\mid H\mid j\rangle
  (a^{(0)}_j + C_1\sum_k\langle j\mid H\mid k\rangle a^{(0)}_k)
\nonumber\\
\frac{a^{(2)}_i}{a^{(0)}_i} &=& 1+(C_1 +C_2)\sum_k
  \langle i\mid H\mid k\rangle\frac{a^{(0)}_k}{a^{(0)}_i}\nonumber\\
  &+& C_1C_2\sum_{jk}\langle i\mid H\mid j\rangle
  \langle j\mid H\mid k\rangle\frac{a^{(0)}_k}{a^{(0)}_i}
\label{e:pl2wgt}
\end{eqnarray}
The $\mid PL2\rangle$ will be closer to the ground state than
$\mid PL1\rangle$, and in principle we can apply powers of $H$
on $\mid PL2\rangle$ to project it toward the ground state.
The PL2-power results should be better than that of PL1-power.
With the similar rough estimation as for PL1 $C_1$, $C_1$ and
$C_2$ in Eq.(\ref{e:pl2wgt}) are near the value $1/E_0$.
Chen and Lee \cite{yctk95} calculated PL2 in 16 sites. The PL2
energy is very close to the ground state energy.
In practice, for larger lattices,
it is quite difficult to find the optimal values
of $C_1$ and $C_2$ by using the VMC method because the behavior
of variational energy in the two-dimensional parameter space
$\{C_1,C_2\}$ is much more complicated than the 1-parameter
case in PL1 calculation. Another problem is that the PL2 calculation
costs much more computer time than PL0 and PL1. We are searching for ways
to carry out the PL2 calculation more efficiently.
In this paper, most data are evaluated from the
PL0 and PL1 calculations.

Note that the energy calculated from the power-Lanczos method
is still a variational estimate and an upper bound of the 
ground state energy. 

Here we demonstrate an example of the energies calculated
by the PL method, and compare that with some recent data
by other groups. For 50 electrons in an $8\times 8$ lattice
and $J/t=0.2$, the PL energies are shown in
TABLE \ref{t:5064}.

It can be seen from TABLE \ref{t:5064} that the PL1 and PL2
energies converge much faster than PL0 (GFMC) ones. 
The upper bound of the ground state energy can be estimated
with less powers such that the fermionic sign problem is
also reduced significantly.

\begin{figure}[htb]
\epsfysize=10cm
\epsfbox{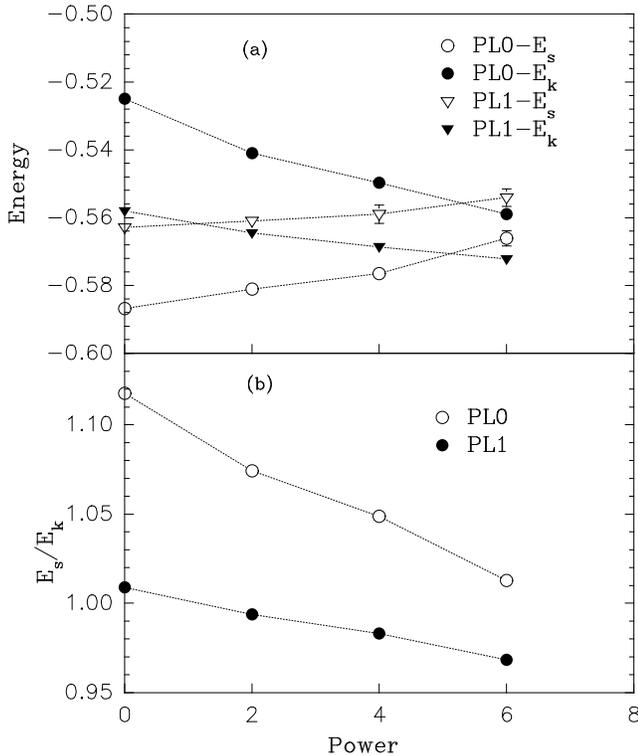}
\caption{(a) $E_s$, $E_k$ and (b) $E_s/E_k$ for different 
powers in $50/64$, $J/t=0.2$.}
\label{f:seke}
\end{figure}

Another interesting point is the ratio of spin ($E_s$) and kinetic
($E_k$) energies. Fig.\ref{f:seke} shows the two portions
of energy of different powers for $50/64$, $J/t=0.2$. It can be
seen that $E_s$ is overestimated while $E_k$ is underestimated
in the trial wave function. From our experiences,
$E_s$ is usually overestimated for the sake of minimizing
the energy of the trial wave function in the physical interested
region of the parameter space. The choice causes misjudgements
of other physical quantities. For example, the superconducting
long range pairing correlation function 
will be strongly overestimated if we choose
the energy-optimized RVB wave function\cite{shih98b}.
From the behavior of $E_s$ and $E_k$, we see that it is a very
important task to find a new trial wave function having more
reasonable $E_s/E_k$ value, as well as having lower trial energy.

In conclusion of this section, we proposed a powerful method
which is a combination of Lanczos iteration and power method.
The Lanczos iteration is used to improve the starting point
of the power method which projects the new (PL) trial wave
function toward the ground state. The ground state properties
can be estimated more accurately under the limitation of the
fermionic sign problem.

\begin{table}[htb]
\caption{Power-Lanczos energies of $J/t=0.2$ for 50 electrons
in a $8\times 8$ lattice.}
\begin{tabular}{|c|c|c|c|}
power & PL0 & PL1 & PL2 \\
\hline
0 & -0.6443(1) & -0.6709(1) & -0.6819(11)\\
\hline
2 & -0.6573(4) & -0.6768(2) & -0.6833(11)\\
\hline
4 & -0.6657(6) & -0.6803(6) & -0.6851(29)\\
\hline
6 & -0.6735(9) & -0.6820(8) & \\
\hline
8 & -0.6769(12) & -0.6855(13) & \\
\end{tabular}
\label{t:5064}
\end{table}

\section{Extrapolation of the Energies}
\label{s:extra}

From TABLE \ref{t:5064} we conclude that the ground state energy of
this system is lower than $-0.6855$ by the variational principle.
HM\cite{hellberg00} gives $-0.6825(23)$
by extrapolating the GFMC data. They used a different method
to calculate energies of all different powers in the same time.
The energies of different
powers calculated by HM are close to our PL0-power
data. It can be seen that the extrapolated GFMC energy is still
higher than the variational upper bound determined by
PL1$_{power=8}$ or PL2$_{power=4}$ from TABLE \ref{t:5064}. 
Our energy vs. power curves for PL0 and PL1 are fitted by
an exponential function:
\begin{equation}
E=E_0+a e^{-bp}
\label{e:fit}
\end{equation}
\begin{figure}[htb]
\epsfysize=7cm
\epsfbox{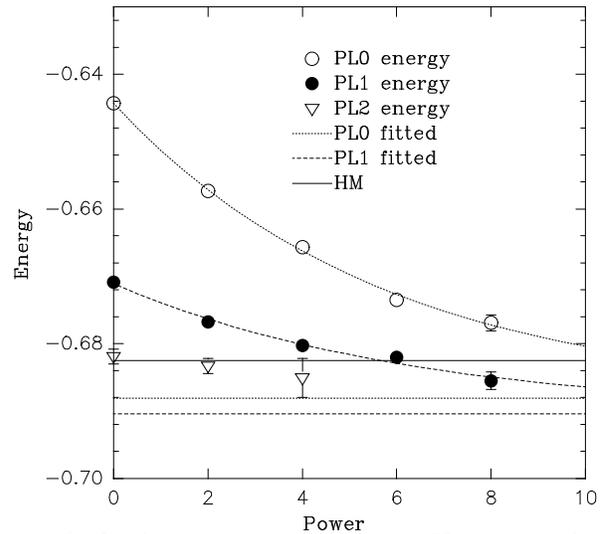}
\caption{Power-Lanczos energies of 50 electrons in $8\times 8$
lattice, $J/t=0.2$. The PL0 (PL1, PL2) data are represented by
open circles (full circles, open triangles). The dotted and dashed
curves fit the PL0 and PL1 data, respectively,
and the horizontal dotted and dashed lines are the extrapolated
values of $power=\infty$. The horizontal solid line is the
ground state value estimated by HM.}
\label{f:5064j2}
\end{figure}
The results are shown in Fig.\ref{f:5064j2}.
For PL0 data, the fitting parameters are $E_0=-0.6881(35)$,
$a=0.0438(33)$, and $a=0.174(26)$. And $E_0=-0.6904(48)$,
$a=0.0193(46)$, and $b=0.158(70)$ for PL1 energies. The most
important parameter is $E_0$, the values for PL0 and PL1
data are consistent, which are both much smaller than
HM result. Since the ground state energy {\it
must be lower than the lowest power-Lanczos energy},
their extrapolated data is clearly inaccurate.

In TABLE \ref{t:j02}, we show $J/t=0.2$ energies for several
electron densities. The lattices and boundary conditions are chosen 
exactly the same as those chosen by Hellberg {\it et al.}\cite{hellberg00}.

$E_{extrap}$ are the data extrapolated from GFMC and
$E_{lanc}$ are the quoted values for Lanczos iterations\cite{hellberg00}.
It can be seen that for this $J/t$, $E_{PL}$, the PL energies
are lower than $E_{extrap}$ for higher densities, and
higher than $E_{extrap}$ for intermediate ones ($50/90$ and
$50/72$).

We emphasize here that
in the cases of high density and small coupling constant, it is
difficult to get accurate enough energies for powers larger than
8 because the sign problem enlarges the statistical
error bar for such large powers. 
Thus it is unreliable to extrapolate the GFMC data
to estimate the ground energy in this regime for larger
clusters, even though the method works well in the
small clusters like $4\times 4$, or at
low density and large $J/t$ regime because the energies
can be calculated very accurately for more powers, and
the convergence in these cases is much faster than
the high density, small $J/t$ in large clusters. For example,
the difference between exact energy and the number evaluated
by PL1-power=10 for 10/16 is about only 0.1\%.

\begin{table}[hbt]
\caption{Energies of $J/t=0.2$ for several densities and
lattices evaluated by PL method $E_{PL}$, extrapolated
GFMC $E_{extrap}$, and Lanczos iterations $E_{lanc}$.}
\begin{tabular}{|c|c|c|c|}
$N_e/N_s$ & $E_{PL}$ & $E_{extrap}$ & $E_{lanc}$ \\
\hline
50/90 & -0.9184(13) & -0.9246(35) & -0.9211(54)\\
\hline
50/72 & -0.8074(11) & -0.8103(28) & -0.8098(32)\\
\hline
50/64 & -0.6855(13) & -0.6825(23) & -0.6826(27)\\
\hline
50/56 & -0.4708(8) & -0.4681(18) & -0.4717(23)\\
\hline
60/64 & -0.3710(6) & -0.3701(17) & -0.3693(7)\\
\end{tabular}
\label{t:j02}
\end{table}

\section{Determination of Phase Separation: Maxwell Construction}
\label{s:maxwell}

There are several criteria to determine the boundary of PS.
The standard one is to find the density of
divergence of the compressibility (the second derivative of
the energy-density curve). This method is successful in
determining the phase boundary of the one-dimensional $t-J$
model, which phase separates into one electron-free region
and another one containing both electrons and holes\cite{hellberg93}.

For the 2D $t-J$ model, the PS state contains one hole-free
Heisenberg AF region and another hole-containing
one. The finite size effect from the ``surface energy'' 
between the hole-free Heisenberg region
of the PS state is significant (of the order $1/\sqrt{N_s}$,
where $N_s$ is the size of the lattice), 
and this surface energy will 
push the PS state to much larger $J/t$ value. If we use
the inverse compressibility method in the 2D $t-J$ model, 
we will get wrong results due to this effect.
Thus we have to use the Maxwell construction method to find the
PS boundary from the energies of the finite-size, uniform
systems.

The energy approaches the ground state as the power increases
in our PL method. Although in the physical regime
most of our best data have not yet converged to the exact ground state
values, the systematic variation of energies is enough for us 
to give a variational estimate of the lower bound of the phase 
separation boundary. We will use the lowest PL energies
in the following discussion.

\begin{figure}[htb]
\epsfysize=7cm
\epsfbox{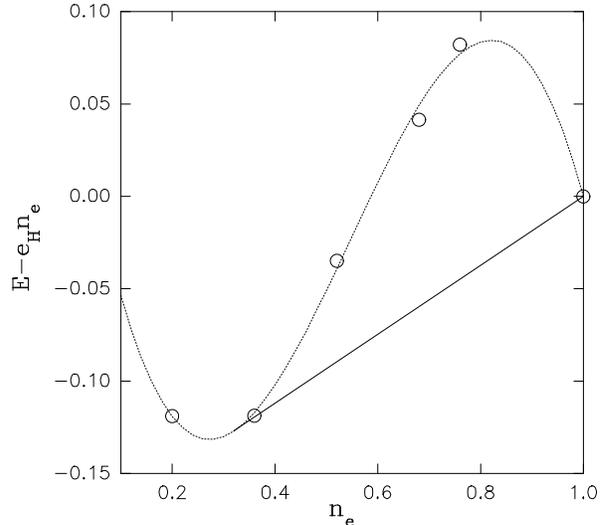}
\caption{The ground state energies of different electron
densities in 50 site lattice for
$J/t=2.5$. The energies are subtracted by the perfect PS
energy $e_Hn_e$. The straight line shows the ground state
energy of the PS state by Maxwell construction. The onset
of PS state is at $n_e=0.320$.}
\label{f:50j25}
\end{figure}

Fig.\ref{f:50j25} shows the energy-density curve of $J/t=2.5$
in 50 sites. The energies are subtracted by $e_Hn_e$, where
$e_H$ is the Heisenberg energy and $n_e$ the electron density.
$e_Hn_e$ is energy of the ``perfect phase separation state'',
which separates into a hole-free Heisenberg antiferromagnet
and an electron-free vacuum phases. This is the ground state
of the infinite $J/t$ limit if the quantum fluctuation near
the boundary of the two regions is not taken into account.
Thus it is also a good reference state for large $J/t$ cases.

Fig.\ref{f:50j25} shows the PS phase boundary determined by
Maxwell construction. It is clear that the curve become convex 
for the larger densities and
the phase separates into a hole-free Heisenberg antiferromagnet
and a region with density $n_e=0.320$. The energy of the PS
state is the linear combination of the two phases, which is represented
by the solid line. The results are consistent with
Hellberg {\it et al.}\cite{hellberg00}, whose critical
$n_e=0.296$.

\begin{figure}[htb]
\epsfysize=10cm
\epsfbox{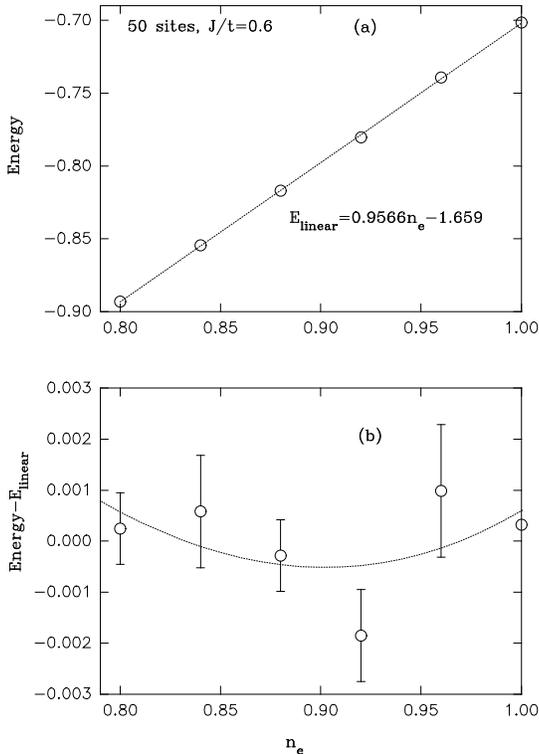}
\caption{
(a) Energy-density curve for $J/t=0.6$ in 50 sites,
it can be fitted very well by a straight line.
(b) Deviation of the energy from the line in (a),
the values are of the same magnitude of the error bars.
}
\label{f:j06}
\end{figure}
For smaller $J/t$, it is more difficult to define the tangent
line to find the critical $n_e$ as we did in Fig.\ref{f:50j25}.
The $J/t=0.6$ data are shown in Fig.\ref{f:j06}(a). It can be
seen that the energy-density curve is almost a straight
line for this coupling strength. The curve is fitted by
a linear function, which is shown by the dotted line in
Fig.\ref{f:j06}(a). And the deviations of the data points
from the line are shown in Fig.\ref{f:j06}(b). We see that
the deviations are of the order of magnitude of the
statistical error bars. Since the subtraction of a linear
function won't change the convexity or concavity of the
curve, the Maxwell construction can be applied on the
curve in Fig.\ref{f:j06}(b) as we did in Fig.\ref{f:50j25}.
A polynomial is used to fit the $E-E_{linear}$ points.
The curve is always concave and there is no PS at this
$J/t$. An interesting point is that one can always find
another fitting function gives PS because the
$E-E_{linear}$ value of $n_e=0.96$ is larger than
those of $n_e=0.92$ and $n_e=1$. Since the $E-E_{linear}$
values are all so small, the fitting is somewhat arbitrary
within the error bars. And the energy-density curve is
almost linear. It is very likely that $J/t=0.6$ is just
the PS phase boundary of vanishing hole density for 50 sites.

Another possibility is the formation of stripe state
suggested by S. R. White {\it et al.}\cite{white00}.
For the stripe state, the holes adjust themselves to form
stripes to minimize the total energy and does not have 
the problem of the surface energy as in the PS state.
The adjustment makes the energy-density curves linear.
But since we use the uniform trial wave functions and
periodic boundary condition here, it is difficult to
get signals of the stripes state. This will be our
future work.

From the two previous examples, we see that
it is difficult to read out the slope variation from the
energy-density plots to do the Maxwell construction, 
especially for small $J/t$ values.
For such $J/t$, the possible critical $n_e$ of PS (if there is)
is close to 1. In this regime, the energy-density curve is very 
close to a straight line and the ground state energies are
very difficult to calculate (or estimate) accurately. It will
be quite unreliable to estimate the critical $n_e$ and
to judge whether there is PS or not by extrapolation and
curve fitting. Therefore we will follow  
Emery {\it et al.}\cite{emery90} by examining a more well-defined
value $e(x)$, the energy per hole for hole density $x$.

The energy of the PS state can be separated into two parts:
\begin{equation}
E\times N_s=(N_s-N)e_H+Ne_h
\label{e:ps1}
\end{equation}
where $N_s$ is the total number of sites and N is the number of sites
in the hole-rich phase. The first term of the right-hand side of
Eq.(\ref{e:ps1}) is the energy of the hole-free Heisenberg region
and the second term is that of the hole-rich part.
$e_H=-1.169J$ denotes the Heisenberg energy
per site \cite{sandvik97}. And $e_h$ is energy
per site in the uniform hole-rich phase, which is a function of
the hole density $x=N_h/N$. $N_h$ is the
number of holes. $E$ can be rearranged into the form:
\begin{equation}
E\times N_s=N_se_H+N_he(x)
\label{e:ps2}
\end{equation}
where
\begin{equation}
e(x)\equiv[-e_H+e_h(x)]/x
\label{e:ps3}
\end{equation}
$e(x)$ can be interpreted as the energy per hole relative to
the half-filled Heisenberg state.
If e(x) of a particular $J/t$ has
a minimum at $x=x_m$ and the hole density of the total system is
smaller than $x_m$, the system will adjust the size of the
hole-rich phase N such that $N_h/N$ is equal to $x_m$ and 
it minimizes the total
energy in Eq.(\ref{e:ps2}). Since $N_s$, $e_H$, and $N_h$ are all
constants, the total energy is minimized as $e(x)$ is minimized. Thus
$x_m$ is the critical density for phase separation at this $J/t$.

We calculated $e(x)$ from the energy of the uniform states $e_h(x)$
by the PL method and found the minimum of $e(x)$ on several sizes of
lattices for several densities and $J/t$. The largest one is the
$\sqrt{122}\times\sqrt{122}$ cluster.
It is very difficult to get the converged ground state
energy in the physical regime due to the sign problem.
But from the systematic PL procedure, we can see the trend of
the change of $e(x)$ and estimate the boundary of the PS state.
The PL-1 power=4 (for 82 and 122 sites) or PL-1 power=6
(for 50 and 36 sites) energy
is used here as the $e_h(x)$. It is about $2\sim4$ percent lower than
the variational energy. We estimate the difference between the
best PL energy is within one or two percentage of the true ground state 
energy.

\begin{figure}[htb]
{\hspace*{1mm}
\epsfysize=6cm\epsfbox{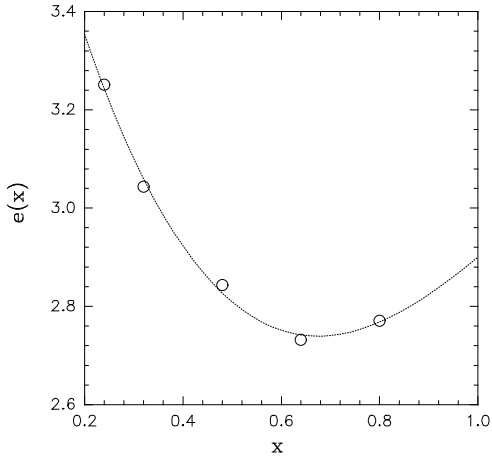}
}
\vskip 4mm
\caption{$e(x)$ vs hole density $x$ of $J/t=2.5$ in 50 sites.
The minimum of the fitted curve is at $x=0.68$.}
\label{f:50ex25}
\end{figure}

In Fig.\ref{f:50ex25} we show the $e(x)$ of $J/t=2.5$ in 50
sites corresponding to Fig.\ref{f:50j25}. The data are fitted
by a polynomial and the minimum of $e(x)$ is at $x=0.68$,
which is exactly agree with the density $n_e=0.32$ determined
from Fig.\ref{f:50j25}. Thus the two versions of 
Maxwell construction are consistent. The advantage of the latter
one is that the critical density for PS state will be determined
much more accurately
if the energy-density curve is almost linear. This is
the case for the physical interested regime. So we will use
the $e(x)$ to determine the PS boundary in this report.

\section{Results and Discussion}
\label{s:result}

\begin{figure}[htb]
\epsfysize=7cm\epsfbox{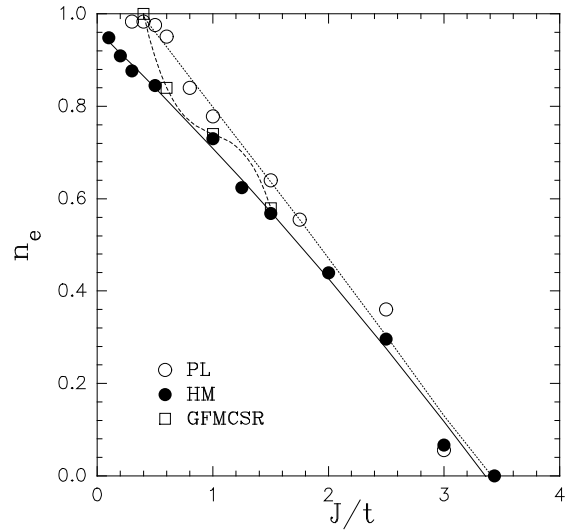}
\vskip 4mm
\caption{Phase separation boundary on the phase diagram of the
two-dimensional $t-J$ model evaluated by the power-Lanczos
method (open circles), GFMCSR method by M. Calandra {\it et al.}
(open squares), and GFMC with extrapolation
by C. S. Hellberg {\it et al.} (full circles).}
\label{f:boundary}
\end{figure}

Using the Maxwell construction we determined the phase boundary
of PS state for several $J/t$. The results are shown in
Fig.\ref{f:boundary}. The results evaluated by using Green
function Monte Carlo with stochastic reconfiguration
(GFMCSR) by M. Calandra {\it et al.}\cite{calandra98} and
Green function Monte Carlo with extrapolation by
HM\cite{hellberg00} are also
shown in the figure.

From Fig.\ref{f:boundary} we see that in the physical
regime, our data are consistent with the GFMCSR results 
that there is a lower bound of the $J/t$ value for PS
state. This is also the case in our previous study\cite{shih98}.
In contrast, the extrapolated GFMC results
show that PS state occurs at all values of $J/t$.
In the large $J/t$ and low electron density region, all the
results are qualitatively consistent.

\begin{figure}[htb]
\epsfysize=7cm\epsfbox{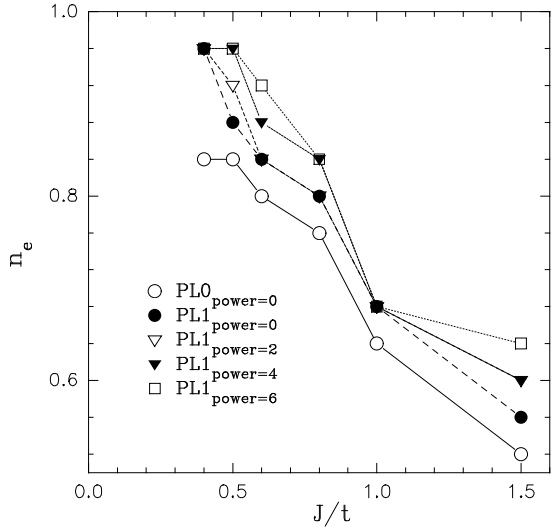}
\vskip 4mm
\caption{PS phase boundaries determined by different PL
levels. It can be seen that for larger powers, the boundary
is pushed to larger $J/t$ and electron density $n_e$.}
\label{f:50pw}
\end{figure}

There are several possible reasons for the inconsistence in
the physical region. First, the energies calculated are
not accurate enough. The energy-density curve is very close
to a straight line in this region, so a small error in
energy will cause a completely different result of Maxwell
construction because the method is related to the second
derivative of the curve. Fig.\ref{f:50pw} shows the PS
phase boundaries determined by energies evaluated by
different projection levels. It is clear that as
the powers increase (that is, the energies closer to the ground
state energies), the PS phase boundary will be pushed to
larger $J/t$ value and electron densities. This feature
appears in all of our calculation on different lattices.
It is strongly suggestive that the PS phase boundary determined
from our variational calculation is a {\it lower bound} of
$J/t$ and electron densities of the real boundary.

\begin{figure}[htb]
\epsfysize=14cm\epsfbox{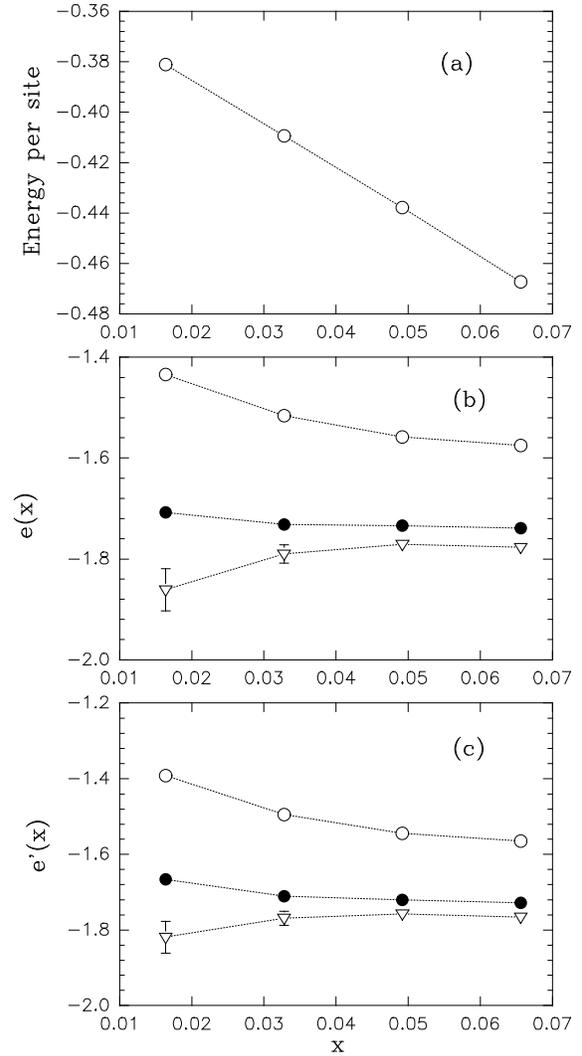}
\vskip 4mm
\caption{(a)Energy vs. hole density $x$ for $J/t=0.3$ in
122 sites. (b)PL0$_{power=0}$ (open circle), PL1$_{power=0}$ (full circle),
and PL1$_{power=4}$ (open triangle) $e(x)$ calculated by using $e_H=-1.169J$
and (c)$e_H=-1.1713J$.}
\label{f:122j03}
\end{figure}

In TABLE \ref{t:j02} we show that our PL energies are
lower than the extrapolated GFMC values. And since PL
method is a variational one, the true ground state
energies will be even lower. We use $J/t=0.3$ in 122 sites
as an example to demonstrate the effect of the non-converged
energies in more detail. Again the energy-density plot in
Fig.\ref{f:122j03}(a) is very close to a straight line
and it will be dangerous to fit the curve and draw the
tangent line from the half-filled point. 

In Fig.\ref{f:122j03}(b) we
show the $e(x)$ vs. hole density $x$ curves for different
level of projection. As the power increases, the energy
decreases and approaches to the ground state energy,
and $e(x)$ also decreases. It is interesting that $e(x)$
of smaller $x$ drops much faster than those of large $x$.
In this case, at the PL0$_{power=0}$ level, the minimal
$e(x)$ will occur at $x$ larger than $0.0656$ (8 holes
in 122 sites). While at the PL1$_{power=0}$ level, the
curve becomes almost flat. For PL1$_{power=4}$, the
smallest density $x=0.0164$ (2 holes) becomes the lowest
one. Note that the energies have not yet converged.
From the trend of $e(x)$ with increasing power, we expect
the minimum of $e(x)$ at $x=0.0164$ will be even deeper. That
is, the PS state occurs at $x\le 0.0164$, while a much
larger value $x_{PS}=0.123$ is reported by HM.

\begin{figure}[htb]
\epsfysize=12cm\epsfbox{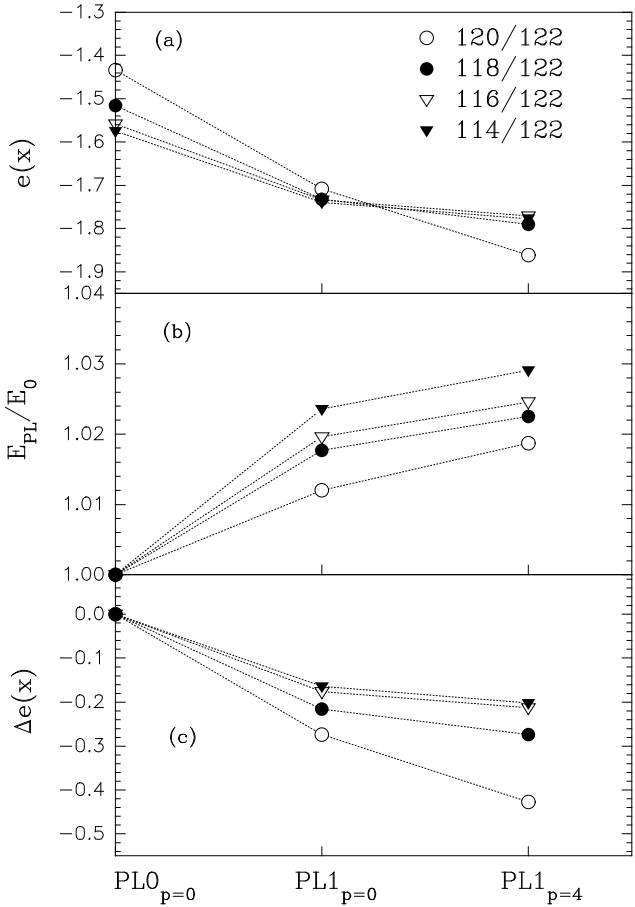}
\vskip 4mm
\caption{(a) e(x) of $J/t=0.3$ for different powers in 122 sites. 
The electron numbers are 120 (open circles), 118 (full circles),
116 (open triangles), and 114 (full triangles). (b)The ratios of
power energies $E_{PL}$ and variational ($PL0_{power=0}$)
energy $E_0$. (c)The differences
between $e(x)$ of different powers and the variational value.}
\label{f:power_e}
\end{figure}

The changes of energies and $e(x)$ with respect to powers
are shown in Fig.\ref{f:power_e}.
It is clear that $e(x)$ drops faster for at lower hole
densities with increasing powers. 
The reason is that the hole density is at the
denominator of the definition of $e(x)$.
From the curves in Fig.\ref{f:power_e}, it is clear
that the energies have not converged yet. And the
trend of the curves strongly suggests
that the $e(x)$ of the smaller $x$ cases will be much more reduced than
those of the larger ones. That is, if the ground state energies
are calculated more accurately, the PS boundary will be 
pushed toward the higher density or larger $J/t$.
Thus the true phase boundary will be at the upper right side
of our variational result in Fig.\ref{f:boundary}.

\begin{figure}[htb]
\epsfysize=14cm\epsfbox{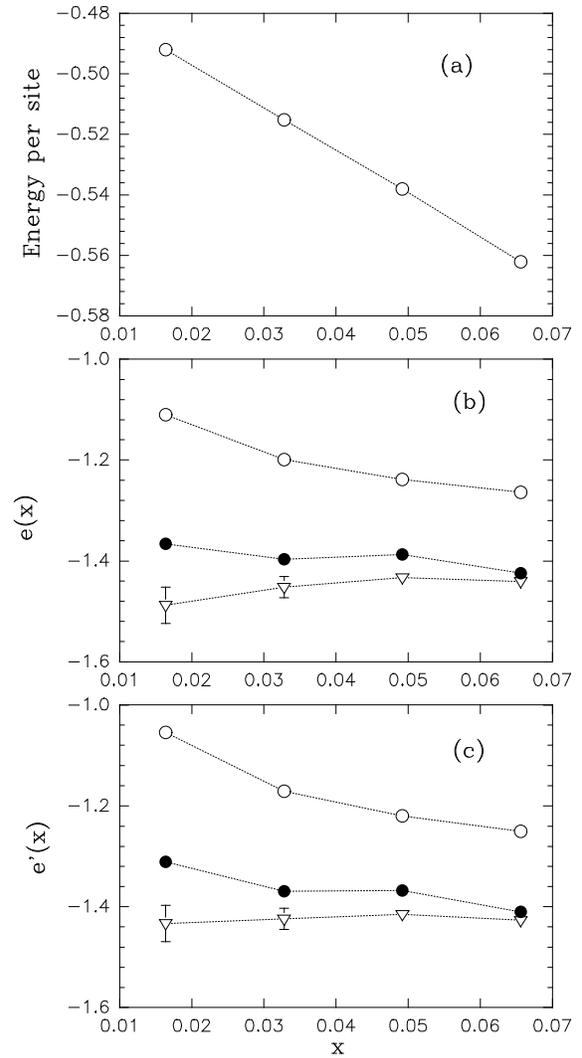}
\vskip 4mm
\caption{(a)Energy vs. hole density $x$ for $J/t=0.4$ in
122 sites. (b)PL0$_{power=0}$ (open circle), PL1$_{power=0}$ (full circle),
and PL1$_{power=4}$ (open triangle) $e(x)$ calculated by using $e_H=-1.169J$
and (c)$e_H=-1.1713J$.}
\label{f:122j04}
\end{figure}

Fig.\ref{f:122j04} shows the similar plots for $J/t=0.4$.
The behavior is a little subtle in this value of $J/t$.
The $e(x)$ is flatter than the $J/t=0.3$ case. The minimal
value is also at $x=0.0164$ point, but within error bar
with the $x=0.0328$ (4 holes) point. Again the energies
are still well above the ground state values, we expect
the $e(x)$ for $x=0.0164$ will be even lower.

Note that in Fig.\ref{f:122j03}(b) and Fig\ref{f:122j04}(b)
the Heisenberg energy used is the thermodynamic limit
$e_H=-1.169J$. And $e_H=-1.1713J$ for finite system with
122 sites are used for Fig.\ref{f:122j03}(c) and Fig\ref{f:122j04}(c).
The later case will make the $e(x)$ values larger. And
in the $J/t=0.4$ case, the PL1$_{power=4}$ curve becomes
almost flat. This finite-size effect will shift the phase
boundary a little but the main results will be unchanged.

Another way to understand the finite-size effect is to use finite-size scaling.
For example, in the case of $J/t=0.4$ in 98 sites shown in Fig.20
in \cite{hellberg00} (calculated by M. Calandra {\it et al.}),
the minimum of $e(x)$ is at the $x_m=0.041$ (4 holes) point.
It is possible that PS occurs at this hole density. But together
with the 50 site data in the same figure, the $x_m=0.08$, also
for 4 holes. The $x_m$ becomes a half as the size of lattice is doubled.
This scaling behavior is consistent with that reported in our
previous paper \cite{shih98}. And the scaling behavior to even larger
lattices is also shown in Fig.3 of Ref.\cite{calandra98}.
Note that all the energies used in these references are upper
bounds of the ground state energies. We expect that $e(x)$ for
smaller hole density $x$ will become even lower, and the phase
boundary of PS state will be pushed to larger electron density
and larger $J/t$.
Thus in contrast with the conclusion by in Ref.\cite{hellberg00},
we conclude that PS does not occur at $J/t=0.4$ for any density.

Another finite size effect comes from the different shell
for different electron density. The effect makes the energy-density
curves jagged and difficult to analyze.
HM calculated the energies for the same
number of electrons (with the same filled shell) in different
sizes of lattices and different boundary conditions to avoid
this effect. But there are still other finite size effects
in this way due to the different sizes and boundary conditions.
The only way to eliminate the finite size effects is to study
this problem in different, and large enough sizes of lattices.

For $J/t\leq0.4$, the minimal $e(x)$ is always at two holes
for lattices of different sizes (up to 122).
This may be resulted from
the two-hole bound state\cite{dagotto92} but not PS at
$x_m=2/N_s$. If there were PS, $x_m$ would be
at the same (or nearby) density  rather than the
same number of holes. 

In conclusion, from the analysis of the energies evaluated
from the PL method, we conclude that $J/t=0.4$ is a lower 
bound for PS. The phases will not separate for $J/t$ smaller than
this value. From the trend of the energies and $e(x)$ functions,
this lower bound may be pushed to even larger $J/t$.
The extrapolation of ground state energies and energy-density
curves may be unreliable and misleading in the physical
interesting regime. F. Becca
{\it et al.} give upper and lower bounds for the ground
state energy of the infinite-$U$ 2D Hubbard model, which is
equivalent to the $J/t=0$ case for the $t-J$ model. Their
analysis on this case ruled out the possibility of PS
for the electron density less than half-filled\cite{becca00b}.
By the way, from the finite size analysis of the exact diagonalization
data for one and two holes in smaller clusters (up to 32 sites)
gives the results that there is no two-hole bound state for
$J/t\leq 0.6$\cite{shih98b,poilblanc93,leung98}. Thus
it seems the attractive interaction is not strong enough to
be the mechanism to cause the PS state for small $J/t$ values.

We thank M. Calandra, S. Sorella, and P. W. Leung for their
kindness to give us their data and very useful discussions.
This work is supported by the National Science Council of
Republic of China, Grants No. NSC89-2112-M-001-103,
NSC89-2112-M-029-004, and NSC89-2112-M-321-001.
Part of the computations were performed on the IBM SP2 Power3
and PC Linux cluster of the National Center
for High-Performance Computing in Taiwan. We are grateful for
their support.

\end{document}